\documentstyle[12pt]{article}
\begin{document}
\def\om{\omega}
\def\omt{\tilde{\omega}}
\def\ti{\tilde}
\def\o{\Omega}
\def\t{T^*M}
\def\vt{\tilde{v}}
\def\ot{\tilde{\Omega}}
\def\otwo{\omt \wedge \om}
\def\owot{\om \wedge \omt}
\def\w{\wedge}
\def\mt{\tilde{M}}

\def\om{\omega}
\def\omt{\tilde{\omega}}
\def\ss{\subset}

\def\om{\omega}
\def\omt{\tilde{\omega}}
\def\ti{\tilde}
\def\o{\Omega}
\def\t{T^*M}
\def\vt{\tilde{v}}
\def\ot{\tilde{\Omega}}
\def\otwo{\omt \wedge \om}
\def\owot{\om \wedge \omt}
\def\w{\wedge}
\def\mt{\tilde{M}}

\def\om{\omega}
\def\omt{\tilde{\omega}}
\def\ss{\subset}
\def\tpm{T_{P} ^* M}
\def\al{\alpha}
\def\alt{\tilde{\alpha}}
\def\la{\langle}
\def\ra{\rangle}
\def\inop{{\int}^{P}_{P_{0}}{\om}}
\def\th{\theta}
\def\tht{\tilde{\theta}}
\def\inox{{\int}^{X}{\om}}
\def\inotx{{\int}^{X}{\omt}}
\def\st{\tilde{S}}
\def\ls{\lambda_{\sigma}}
\def\p{{\bf{p}}}
\def\pb{{\p}_{b}(t,u)}
\def\pbm{{\p}_{b}}
\def\d{\partial}
\def\d+{\partial_+}
\def\d-{\partial_-}
\def\pat{\partial_{\tau}}
\def\pas{\partial_{\sigma}}
\def\dpm{\partial_{\pm}}
\def\l2{\Lambda^2}
\def\be{\begin{equation}}
\def\ee{\end{equation}}
\def\bea{\begin{eqnarray}}
\def\eea{\end{eqnarray}}
\def\ej{{\bf E}}
\def\ed{{\bf E}^\perp}
\def\si{\sigma}
\def\cg{{\cal G}}
\def\cgt{\ti{\cal G}}
\def\cd{{\cal D}}
\def\ce{{\cal E}}
\def\bz{\bar{z}}
\def\e{\varepsilon}
\def\b{\beta}
\begin{titlepage}
\begin{flushright}
{}~
\end{flushright}

\vspace{1cm}
\begin{center}
{\Large \bf  Poisson-Lie  T-duality:}\\
{\Large \bf  Open Strings and  $D$-branes}\\[20pt]
{\small
{\bf C. Klim\v{c}\'{\i}k}\\
Theory Division CERN \\
CH-1211 Geneva 23, Switzerland \\[10pt]
{\bf P. \v Severa }\\
Department of Theoretical Physics, Charles University, \\
V Hole\v sovi\v ck\'ach 2, CZ-18000 Praha,
Czech Republic\\[20pt] }

\begin{abstract}

Global issues of the Poisson-Lie T-duality are addressed. It is shown
that oriented open strings propagating on a group manifold $G$ are dual to
   $D$-brane - anti-$D$-brane pairs
propagating on the dual group manifold $\ti G$. The $D$-branes coincide
with  the symplectic leaves of the standard Poisson structure induced on
the dual
group $\ti G$ by the dressing action of the group $G$.
T-duality maps the momentum
of the open string into the mutual distance of the $D$-branes in  the pair.
The whole picture is then extended to the full modular space $M(D)$
of the Poisson-Lie
equivalent $\si$-models which is the space
 of all Manin triples of a given Drinfeld double.
 T-duality rotates the zero modes of  pairs of
 $D$-branes living on  all group targets belonging to $M(D)$.
In this more general case the $D$-branes are
 preimages of symplectic leaves
in certain Poisson homogeneous spaces of their targets and, as such,
they are either all even or  all odd dimensional.
\end{abstract}
\end{center}
\vskip 0.5cm

\end{titlepage}

\section{Introduction}

This is our first note where we address the global issues of the Poisson-Lie
T-duality \cite{KS2}.
We believe that there is a little doubt that Poisson-Lie
T-duality  does naturally generalize the Abelian  \cite{SS}
--\cite{Alv1} and the traditional non-Abelian \cite{FJ}--\cite{GRV}
T-dualities. After the original work \cite{KS2}, there was a further
development on the subject \cite{T,AKT,TU,KS3} where there was demonstrated
that the Poisson-Lie T-duality enjoys most of the basic characteristic
features of the Abelian T-duality at both  classical and quantum  level.
However,
in order to complete the full analogy between the standard Abelian
and Poisson-Lie T-duality it is crucial to understand the issue of the
zero modes or, in other words, the global issues of the Poisson-Lie
T-duality.  Already in our first paper on the subject
dealing with closed strings, we have proved
 the classical phase space equivalence of the  mutually dual sigma-models
only for restricted phase spaces deprived of zero modes. Actually we
had to restrict the phase space of closed strings of the $\si$-model on a group
manifold $G$  by the constraint of unit monodromy with respect to the
dual group and vice versa (for details see \cite{KS2}). For instance,
in the case of
the Abelian
Drinfeld double (standard Abelian T-duality) the both mutually dual
unit monodromy constraints eliminate all momentum  and
winding zero modes and leave just the oscillator modes.  Hence, only
the local aspects of T-duality can be recovered in this way.
Only in the case of the standard Abelian duality,
and remarkably
 at the quantum level, things become better and for a specific
choice of the geometry of the double (the adjustment $R$ and $\alpha'/R$
of the lengths of the dual homology cycles) the duality extends to the
zero modes in the standard way \cite{SS,RoVe}.

Already for the case of the traditional non-Abelian duality
\cite{FJ,FT,OQ,GR,GRV}
 the  understanding
of the global
issues concerning closed strings is  generally considered as a difficult
problem. A constructive statement was given e.g. in \cite{GR}
where the authors claimed that the dual CFT is the $G$ orbifold of the
original theory and $G$ is the non-Abelian part of the (semi-Abelian)
Drinfeld double governing the traditional non-Abelian duality \cite{KS2}.
We are also trying elsewhere to understand the global issues of the Poisson-Lie
T-duality for closed strings, however, our hope is to show that Poisson-Lie
T-duality is the  symmetry of a {\it single} CFT \cite{KS4}.

Rather surprisingly, we find the global issues of T-duality much easier
to treat in the case of open strings. The standard Abelian T-duality
for  open strings is now a hot topic.
 The subject originated in the papers
\cite{HoSa,Gre,Pol,PoWi} were the appearance of the Dirichlet boundary
conditions (hence $D$-branes) was understood.
 In this note we shall show that a similar picture emerges in the Poisson-Lie
case.
The momentum zero modes  of  open strings are mapped into distance  zero modes
of pair of $D$-branes propagating on the dual group manifold and vice-versa.
The $D$-branes propagate as if there were equally big charges
with  the opposite signs at the ends of the
attached Dirichlet strings.  These charges feel
the field strengths on the $D$-branes given by the symplectic forms on them.
The details we give in section 2.

It may seem that such a T-duality relates objects of different kinds.
 In fact,  we shall show that the T-duality between open strings and pairs of
$D$-branes is just the limiting case of a more general duality between pairs
$D$-branes on $G$ and pairs of $D$-branes on $\ti G$ group targets.
In section 3, we shall describe
the construction, which heavily use the results of Dazord and Sondaz \cite{DS}
and Lu \cite{Lu} on Poisson structures on $G$  compatible with the action
of the Poisson-Lie group $G$ on itself and the results of Drinfeld \cite{D2}
on Poisson homogeneous spaces of Poisson-Lie groups.

\section{Open strings - $D$-branes duality}

Consider a $\si$-model Lagrangian on a group manifold $G$

\be L=(R+\Pi(g))^{-1}(\partial_+ g g^{-1}, \partial_- g g^{-1}),\ee
where the indices $\pm$ means the light cone variables on the world
sheet and  $R$ is a nondegenerate bilinear form
(with also a  non-degenerate symmetric part)  on the dual space $\cg^*$ of
the Lie algebra $\cg$ of the group $G$. $\Pi(g)$ is such  a bivector field
on the group manifold  which  gives a Poisson-Lie bracket on $G$
(i.e. the multiplication $G\times G\to G$ is the Poisson map).         The
 model (1) then has a Poisson-Lie symmetry with respect to the right
action of     $G$ on itself \cite{KS2}\footnote{This Poisson-Lie symmetric
model with respect to the right action of $G$ was first constructed in
\cite{KS2}. However, in distinction to the present case, in that original
 paper we used the {\it left}-invariant
currents instead of the {\it right}-invariant currents  and the $\si$-model
(1) was therefore written in \cite{KS2} in more cumbersome way. By starting
from \cite{KS3}, we use the right-invariant currents and an interested
reader may choose \cite{KS3} as a reference background for reading this note.}
and the Poisson bracket $\Pi(g)$ in the
standard way furnishes the coalgebra $\cg^*$ with the Lie algebra structure
(for a review see \cite{D}). We denote as $\cgt$ the coalgebra $\cg^*$ with
this new  Lie  algebra structure and $\ti G$ the corresponding group.
There is one consistency requirement on  the structure constants
of the both Lie algebras \cite{D} which is invariant upon the exchanging
the algebras. Hence,
 there is a beautiful duality between the groups $G$ and $\ti G$
discovered by Drinfeld \cite{D}. There is a Poisson-Lie bracket
$\ti \Pi(\ti g)$ on the dual group manifold $\ti G$ such that it
converts the coalgebra $\cgt^*$ precisely into the original Lie algebra $\cg$.
In \cite{KS2} we have argued that the T-duality in string theory is just the
manifestation of this Poisson-Lie duality, because for the closed
strings deprived of zero modes we have shown that the $\si$-model (1)
is equivalent to the following $\si$-model on the dual group $\ti G$ manifold:
\be \ti L=(R^{-1}+\ti \Pi(\ti g))^{-1}(\partial_+ \ti g \ti g^{-1},
\partial_- \ti g \ti g^{-1}).\ee
The equivalence of the two $\si$-models at the classical level means the
existence of the canonical transformation between the phase spaces of the
models which preserves the Hamiltonians.

 We have  shown in \cite{KS3} that the phase space of the mutually
dual models  coincides
with the loop group $LD$ of the Drinfeld double $D$. The Drinfeld double $D$ is
a group whose Lie algebra as a vector space is the direct sum of the
vector spaces $\cg$ and $\cgt$. The commutators within $\cg$ and $\cgt$ do not
change and commutators between $\cg$ and $\cgt$ are given by the combination
of the coadjoint actions of $\cg$ on $\cg^*=\cgt$ and $\cgt$ on $\cgt^*=\cg$
\cite{D}. The standard pairing $\la .,. \ra$
 between the algebra $\cg$ and its coalgebra
$\cgt$ is then interpreted as an invariant non-degenerate bilinear form
on ${\cal D}$ such that $\la \cg,\cg \ra = \la \cgt,\cgt \ra =0$
\footnote{There exists a different but equivalent way how to construct
the bialgebra $(\cg,\cgt)$ by starting from a Lie group $D$ (the
 Drinfeld double)  such that
its Lie algebra
$\cd$, viewed as the linear space, can be
 decomposed into
a direct sum of vector spaces which are
 themselves
 maximally isotropic subalgebras with
respect to a non-degenerate invariant
bilinear form on $\cd$ \cite{D}.
 An isotropic subspace of $\cd$ is such
 that the value of the invariant form
on any
two vectors belonging to the subspace
 vanishes (maximally isotropic means
that this subspace cannot be enlarged
 while preserving its  isotropy).
 Any such decomposition of the double
 into
a pair of maximally
isotropic subalgebras
$\cg + \cgt=\cd $ is usually referred to
as the Manin triple and the pair $(\cg,\cgt)$ as the Lie bialgebra.}.
 In \cite{KS3} we have also written a duality invariant first order
(Hamiltonian) action on $LD$ which, upon choosing different parametrizations
of $LD$ and solving away `halves' of fields, yields both $\si$-models
(1) and (2) of the dual pair for the case of the closed strings deprived
of zero modes. The explicit form of the Lagrangian of
this action is as follows
($l(\si,\tau)\in D$)

\be {\cal L}={1\over 2}\la \pas l~l^{-1},\pat l~l^{-1}\ra+{1\over 12}
d^{-1}\la dl~l^{-1},[dl~l^{-1},
dl~l^{-1}]\ra +{1\over 2}\la \pas l~l^{-1}, A\pas l~l^{-1}\ra\ee
Here $\la .,.\ra $ denotes the non-degenerate invariant bilinear form
on the Lie algebra ${\cal D}$  of the double. In the second term
 in the r.h.s. we
recognize
the two-form
potential
of the WZW three-form on the double and
$A$ is a linear (idempotent) map from the Lie algebra ${\cal D}$
of the double into itself. It has two eigenvalues $+1$ and $-1$, the
corresponding eigenspaces ${\cal R}_+$ and ${\cal R}_-$
have the same dimension $dimG$, they are perpendicular to each other
in the sence of the invariant form on the double and
they are given by the following recipe:
\be {\cal R}_+=Span\{t+R(t,.),t\in\cgt\},\quad  {\cal R}_-=
Span\{t-R(.,t),t\in\cgt\}.\ee
Thus the modular space of such actions is described by (non-degenerate)
bilinear forms $R(.,.)$ (matrices) on the algebra $\cgt$ \footnote{Since
$R(.,.)$ is non-degenerate, there exists
 the inverse bilinear form defined on the dual $\cg$ of
$\cgt$, hence such description of the modular space does not break the
duality.}.
For a better orientation of an interested reader we stress that the
first two terms in (3) give together the standard WZW Lagrangian on the double
if we interpret  $\tau$ and $\sigma$ as  the `light-cone' variables. These two
first terms
play the role of the `polarization'  term $pdq$ in the first order
variational principle
$$ S=\int L=\int pdq-Hdt.$$
The remaining third  term of the action (3) plays the role of the Hamiltonian
$H$. The field equations coming from (3) are very simple:
\be \la \partial_{\pm} ll^{-1},{\cal R}_{\mp}\ra=0.\ee

 As we have mentioned in the Introduction, the  inclusion of the
zero modes for the closed strings is a difficult problem. Instead, let us
consider
an oriented {\it open} string whose propagation on a group manifold $G$
is governed by the $\si$-model (1). The boundary conditions at the end-points
of the open string are standard: there should be no  flow of momentum
through the end-points. However, we should be more careful in understanding
what the `momentum' means in this case where the background is not
isometric. In our case (1) the Noether currents corresponding to the
right action of the group $G$ on itself are not conserved on shell (because the
dependence on $g$ of the Poisson bracket $\Pi(g)$ spoils the right invariance
of the action). The Noether current one-form $J(g)$ is  defined by
the variation
of the action (1)
\be \delta \int L =\int \la J(g) \stackrel {\wedge}{,} d\epsilon \ra
+\int \epsilon^a {\cal L}_{v_a}(L) \ee
where $g+\delta g=g(1+\epsilon)$, $\epsilon\in\cg$ and
${\cal L}_{v_a}$ are the Lie derivatives of the Lagrangian with respect
to the left invariant vector fields on $G$. Clearly, the
(world-sheet) one-form $J(g)$ is an element of the coalgebra $\cg^*$
which itself has the Lie algebra structure $\cgt$. The explicit form
of $J(g)$ in terms of the $\si$-model matrix was given in \cite{KS2} where
it was also shown that for the $\si$-model (1) the equations of motions
can be written as $\cgt$-valued zero-curvature condition for the $\cgt$-valued
`connection' form $J(g)$. In other words, a ($\ti G$-valued) quantity
\be \ti H =P\exp{\int_{\gamma} J(g)}\ee
is conserved.
 Here $P$ means the path-ordered exponential and $\gamma$ is
an arbitrary curve crossing the world-sheet of the open string. Hence, $\ti H$
is our $\ti G$ valued non-commutative momentum and the boundary conditions
are such that the component of the momentum density
one-form $J(g)$ along the boundaries vanishes at the end-points of the
string. Note that the zero-curvature condition on $J(g)$ means that
though $J(g)$ is not a conserved current nevertheless
its Wilson line (7) still gives the (non-commutative) conservation law.

In \cite{KS2} we have argued that every extremal string $g(\si,\tau)$
of the model
(1) propagating on the group manifold $G$   can be naturally understood
as propagating also in the Drinfeld double $D$. The reason is that for
the extremal string the one-form $J(g)$ is the flat $\cgt$-connection therefore
it can be written as
\be J(g)=d\ti h \ti h^{-1}, \quad \ti h\in\ti G.\ee
Hence the string propagation on  the double can be described by  mapping
\be l(\si,\tau)=g(\si,\tau)\ti h(\si,\tau),\ee
where $g$ and $\ti h$ are simply multiplied in the Drinfeld double
sense. Such $l(\si,\tau)$ then fulfils Eqs. (5).
Note that given $J(g)$, the corresponding $\ti h$ is given up to
a constant element $\ti h_0\in \ti G$. This means that the extremal
string $g(\si,\tau)$ is lifted into the double up to a constant right
translation $\ti h_0$ in the double.

What happens to the boundaries of the extremal strings lifted to the double?
Because of the boundary conditions, the end-points moverespectively
 along two copies
of the group manifold $G$, i.e. $G\ti h_i$ and $G\ti h_e$,
 embedded into the double by the right action
of two constant elements $\ti h_i$ and $\ti h_e$ from the dual group $\ti G$.
Those elements are constrained by the equation
\be \ti h_e=\ti h_i \ti H,\ee
where $\ti H$ is the conserved (non-commutative) momentum of the string.
Using our old strategy \cite{KS2}, we may find  projections
$\ti g(\si,\tau)$ into the dual group of the extremal strings $l(\si,\tau)$
living in  the double  according to the relation
\be l(\si,\tau)=\ti g(\si,\tau) h(\si,\tau), \quad h\in G.\ee
Under this projection the manifolds  $G\ti h_i$ and $G\ti h_e$ of the
string end-points in the double get projected
by definition into the so-called dressing orbits or orbits of the dressing
action of the group $G$ on its dual group $\ti G$ \footnote{If the dual group
is commutative, then the dressing action of $G$ is just the co-adjoint
action of $G$ on its coalgebra $\cg^*$.} \cite{D}. Those dressing orbits
coincide with the symplectic leaves of the Poisson-Lie bracket
$\ti \Pi (\ti g)$
on $\ti G$ discussed above.

We already know from our previous works \cite{KS2,KS3} that the dynamics
of the bulk
of the string in the dual group $\ti G$ ( corresponding to the $\si$-model
(1) on $G$) is governed by the $\si$-model (2) on $\ti G$. Now  we have to
care about the boundary conditions. We observe that the standard open string
boundary conditions on $G$ give rise to the  Dirichlet boundary
conditions for the open strings propagating on $\ti G$. The end-points
of the strings stick on the symplectic leaves of the Poisson-Lie bracket
$\ti \Pi (\ti g)$ in $\ti G$. These leaves
 we standartly interpret as the $D$-branes
\cite{Pol}. They are automatically even dimensional
and  the mutual geometry of the $D$-branes corresponding respectively
to the string end-points is given by the momentum $\ti H$ of the open string
in $G$.  As we have  mentioned above by a suitable choice of $\ti h_0$
we are actually free to set the value of $\ti h_i$ to the dual group unit
element $\ti e$. Then the submanifold $G\ti e$ of the double coincides with
the embedding of the group $G$ into the double $D$ and its projection
into $\ti G$ is just one-point symplectic leaf $\ti e$.
The whole picture of duality then becomes particularly transparent: one
of the $D$-branes of the pair is just the  origin $\ti e$ of $\ti G$
and the other
is the symplectic leaf in $\ti G$ to which belongs
 the momentum $\ti H$ of the open string in $G$.

It is obvious, that apart from sticking on the $D$-branes, something more
must hold for the motion of the string end-points {\it within} the $D$-branes.
Indeed they must move in such a way that their dual open strings in $G$
have vanishing momentum flow through their boundaries. Hence we feel
that a boundary term has to be added to the dual model bulk action (2)
 in order to establish the perfect  duality. It is not in fact that difficult
to find this  boundary term. We may use the first order
duality invariant Lagrangian (3) on the double derived in \cite{KS3} but
we have to specify the boundary conditions on the fields $l(\si,\tau)$.
They are such that in the decomposition $l=g\ti h$ of the double the
field $g$ may be arbitrary but $\ti h$ has to be constant along the boundaries
of the worldsheet. With such boundary conditions the action coming from
the Lagrangian (3) is not well defined because for the opens strings
various possible choices of the inverse exterior derivative of the WZW term
are inequivalent. We pick up such a choice of $d^{-1}$ of WZW which guarantees
that the Polyakov-Wiegmann formula \cite{PW} holds\footnote{This choice of
$d^{-1}$ of WZW form is given by the antisymmetric part of the bilinear form
$\la ., \Pi_{\ti LR} .\ra$ defined in \cite{KS2}.}

\bea   \int \alpha&=
          {1\over 2}\int \la \pas l~l^{-1},\pat l~l^{-1}\ra+{1\over 12}
\int d^{-1}\la dl~l^{-1},[dl~l^{-1},
dl~l^{-1}]\ra \cr &= \int \la \pas \ti h\ti h^{-1},g^{-1} \pat g\ra.\eea
This WZW action on the double corresponds to
the polarization form $\alpha(=pdq)$ part  of the first
order action (3).

We may now choose the dual parametrization $l=\ti g h$. Then we know
that at the end-points of the string, $\ti g$ is constrained to
                          some symplectic leaves.            We cannot
expect, however, that the Polyakov-Wiegmann formula holds also for the dual
parametrization $l=\ti g h$ because another choice of $d^{-1}$ of the WZW form
would be needed to ensure this\footnote{The choice of the antisymmetric part
of $\la .,\Pi_{ L\ti R}.\ra$ in the sense of \cite{KS2}.}.
 Fortunately, we have found that those
two choices of $d^{-1}$ of WZW term differ \cite{KS2} by a
non-degenerate  closed (and hence symplectic) two-form
$\Omega$ on the double constructed by Semenov-Tian-Shansky \cite{SM,KS2,KS3}.

Hence  we may write
\be \int \alpha =
\int \la \pas hh^{-1}, \ti g^{-1} \pat \ti g\ra +\int \Omega (\ti gh).\ee
Because $\Omega$ is closed  the boundary terms do contribute
and the bulk $\si$-model action (2) gets supplemented with the boundary
terms
\be S_{b}=\int_{i}d\tau  A_{\mu}(x)\pat x^{\mu}-\int_{e}d\tau
A(y^{\nu})\pat y^{\nu}.\ee
Here $x^{\mu}$ and $y^{\nu}$ are some coordinates on the symplectic leaves,
 $A_{\mu}(x)$ and $A_{\nu}(y)$ are
  electromagnetic potentials on the leaves and the indices $i$ and $e$
denote the end-points of the string. The minus sign between
the boundary contributions means that the end-points of the string
carry equally large but opposite charges. Obviously, the
exact form of the  electromagnetic
potential come from the Semenov-Tian-Shansky form on the double
which induces the standard  (coming from the Poisson-Lie structure)
 symplectic form on the symplectic leaves on $\ti G$. These symplectic
forms on the leaves give just the field strenghts of the potentials
$A_{\mu}$.

We have shown that one of the symplectic leaves on $\ti G$, where the
end-points of the string live, can be chosen to be the group origin $\ti e$.
The opposite end-point then lives on the symplectic leaf which corresponds
to the total momentum $\ti H$ of the dual open string living on $G$
and carry a charge which feels the field of the symplectic form
on the leaf. In some examples this field is nothing but the field
of a monopole sitting at the origin $\ti e$.

\section{$D$-branes - $D$-branes duality}

It may seem that the duality described in the previous section relates
apparently different objects: open strings and $D$-branes. Here we shall
show that such a duality is rather a singular case of a more `symmetrically'
looking duality between $D$-branes and $D$-branes. Indeed, suppose
that our Drinfeld double based on the bialgebra $(\cg,\cgt)$ has
a different decomposition (Manin triple, cf. footnote (6)) into a pair
of maximally isotropic subalgebras ${\cal K}$ and $\ti {\cal K}$. This
means that the Drinfeld doubles based on the bialgebras $(\cg,\cgt)$ and
$({\cal K},\ti{\cal K})$ respectively coincide.
In what follows we shall consider $({\cal K},\ti {\cal K})$
and $(\ti {\cal K},{\cal K})$ as two {\it different} points of the modular
space
$M(D)$ of all Manin triples corresponding to the given Drinfeld double.
Whenever we shall speak about a target from $M(D)$, we shall mean
a group corresponding to the first Lie algebra of the bialgebra
from $M(D)$. Now consider again our
`old' model of open strings on the target $G$ whose dynamics is
governed by the $\si$-model (1).   We already now that from the point
of view of the dual manifold $\ti G$ this model is equivalent to
the $\si$-model (2) supplemented with the boundary terms (14); how
it looks from the point of view of the group manifold $K$ corresponding
to ${\cal K}$? The answer
for the bulk part we already know from \cite{KS2,KS3}; it turns out that
the boundary terms also can be elegantly written in terms of the well-known
structures on the Drinfeld double. So the action on $K$
reads

\be S=\int d\si d\tau (E+\Pi(k))^{-1}(\partial_+ k k^{-1}, \partial_- k k^{-1})
+  \int_{i}  A_{\mu}(x)\pat x^{\mu}-\int_{e}
A_{\nu}(y)\pat y^{\nu} ,\ee
where $E$ is a constant bilinear form on $\ti {\cal K}$ related to the form
$R$ in (1) by a constant projective transformation \cite{KS2,KS3} and
$\Pi(k)$ is the Poisson bracket on $K$ corresponding to $\ti{\cal K}$.
The meaning
of $A$'s will become clear soon and the coordinates $x$ and $y$ parametrize
the projections into $K$ of the
submanifolds $G\ti h_i$ and $G\ti h_e$ of the double.  Recall that those
are submanifolds where the end-points of the string  live. The projection
$k(\si,\tau)$  of the string from the double into $K$ is defined in the
standard way:
\be l(\si,\tau)=k(\si,\tau)\ti m(\si,\tau), \quad \ti m\in\ti K.\ee
Now we use the result of Dazord and Sondaz \cite{DS} and Lu \cite{Lu}
 which
states that all Poisson (but not necessarilly Poisson-Lie!) structures
on the group $K$ compatible with the right action of the Poisson-Lie group
$K$ on itself\footnote{Compatibility means that the multiplication
map $K\times K\to K$ is a Poisson map. Of course, here the first copy
of $K$ and the image $K$ has the Poisson structure in question and the
second copy of $K$ which acts from the right has the Poisson-Lie structure
induced by the dual group $\ti K$.} are in one-to -one correspondence
with the isotropic subalgebras in ${\cal D}$ transversal
to ${\cal K}$ (=having vanishing
intersection with
 ${\cal K}$). Suppose that from the point of view of  ${\cal K}$ the algebra
$\cg$ is indeed such a transversal algebra. Using the results of \cite{DS,Lu}
we can easily find the symplectic leaves of this Poisson-structure
on $K$. One has first to exponentiate the transversal isotropic subalgebra
(in our case $\cg$)
into the corresponding subgroup of the double ($G$), then shift it by
the right action of an arbitrary constant element $r$ from the double $D$
and finally project the submanifold of $D$ obtained in this way into
$K$. Remarkably, it follows that the end-points of our open string in $K$
move just along these symplectic leaves of the Poisson-structure induced
on $K$ by the isotropic algebra $\cg$. Moreover, the potential $A_{\mu}(x)$
has the field strength equal to the symplectic form induced on the leaf
by the Poisson structure on $K$ induced by $\cg$. The mutual geometry
of the symplectic leaves, or $D$-branes, in $K$ is given by the momentum
$\ti H$ of the corresponding open string in the target $G$.

If the algebra $\cg$ is not transversal to ${\cal K}$ then we have to use
a generalization
of the results \cite{DS,Lu} given by Drinfeld \cite{D2}.  The   $D$-brane
structure
on $K$ is, of course, also
 obtained in this case by projecting  the submanifolds
$G\ti h_i$ and $G\ti h_e$ into  $K$.
But now the $D$-branes need not be symplectic leaves of a Poisson structure
on $K$. It turns out, however, that they {\it are} preimages of
the symplectic leaves of certain Poisson komogeneous
$K$-space\footnote{A Poisson homogeneous $K$-space is a homogeneous
$K$-space equipped with a Poisson structure compatible with the action
of the Poisson-Lie group $K$. Drinfeld has shown in \cite{D2} that all
such Poisson homogeneous
spaces correspond to  maximally isotropic subalgebras of
the double.}. This Poisson homogeneous $K$-space is obtained as a
left coset of $K$ by a subgroup  whose Lie algebra is the intersection
of $\cg$ and ${\cal K}$. The Poisson structure on it is induced by $\cg$
in the sense of \cite{D2}.
The pullback into $K$ of the symplectic form on a   symplectic leaf in
this coset $K$-space gives the field stregth  on the pre-image of the
leaf(=$D$-brane)
in $K$.
In particular, if $\cg$ is ${\cal K}$ itself
 then the coset is just one point, its
pre-image and hence $D$-brane is the whole group $G$ and the field strength on
the $D$-brane
trivially vanishes. This is the $D$-brane interpretation of the open strings
in a target $G$. Note that the dimensions of $D$-branes for an arbitrary
target $K$ from the modular space $M(D)$ are either all even or all odd
dimensional. This
follows from the fact that the dimension of the $D$-brane is equal to the
dimension of the corresponding symplectic leaf in the coset plus the
dimension of the intersection of $\cg$ and ${\cal K}$.

  Thus we may say that, in general, the duality
we are describing just rotates $D$-branes on various targets from $M(D)$.
It is clear that the structure of the boundary terms which supplement the
bulk actions of the $\si$-model from $M(D)$ (cf. Eq. (15)) is given
solely by an isotropic subalgebra of the double.
There is an interesting possibility that this isotropic subalgebra ${\cal F}$
 may not belong to $M(D)$ which means that it does not possess
an isotropic dual subalgebra $\ti {\cal F}$. Then our duality would rotate
just `pure' $D$-branes in all targets from $M(D)$ which means that there is
no target in $M(D)$ for which the corresponding
$D$-branes would be equivalent just to open strings. We may therefore conclude
by stating the main result of our note:
\vskip 1pc
\noindent{\it Theorem}: A Poisson-Lie equivalent class of $D$-brane
theories is given by a pair: a Drinfeld double $D$ and a maximally isotropic
subalgebra ${\cal F}$ in it.
For an arbitrary group target $K$ from $M(D)$ the $D$-brane theory
dual to the other theories in $M(D)$ is described by the action of the form
(15) with the
appropriate form $E$ \footnote{$E(.,.)$ is given by  writing the subspaces
${\cal R}_{\pm}$
 in the same way as in (4) but  now with $t\in \ti{\cal K}$.}.
 The $D$-branes coincide with the  preimages of the symplectic leaves
of the Poisson structure
in the left coset of the group $K$ by the subgroup given by the intersection
of ${\cal K}$ and ${\cal F}$. The field strength of the boundary potential
$A_{\mu}$ on the $D$-brane is given by the pullback of the symplectic form
on the symplectic leaf. The mutual geometry of the $D$-branes in the pair
is given by one constant element $d_0$ from the double (more precisely
by an element of the left coset of the double by the group $F$  having the Lie
algebra  ${\cal F}$). This element  measures the `distance'
of the two copies of the group manifold $F$
lifted in the double by right action of two arbitrary constant elements from
the double.

The proof of the Theorem is simple and more or less straightforward.
\vskip1pc
\noindent {\it Remark}: In the case of the Abelian T-duality the Drinfeld
double is a $2d$-dimensional Abelian group and the modular space
$M(D)$ is the coset $O(d,d.R)/O(d,R)\times O(d,R)$ (for the simply connected
double). An arbitrary maximally isotropic subalgebra ${\cal F}$ in the double
does have its dual $\ti {\cal F}$ and, hence, it is an element of $M(D)$.
All corresponding $D$-branes theories are  then equivalent to the open string
theory on the target $F$.
\section{Outlook}

Among easy  open problems which should be addressed we may mention
an understanding of Buscher's duality for $D$-branes
where the $\si$-model manifold
is a product of some manifold $M$ and a group G and the $\si$-model
 is Poisson-Lie symmetric with respect to the right action of the
group $G$ on the target. Another not difficult problem
would be a path integral argument which would provide a
quantization of the described global picture of $D$-branes duality.
We shall probably solve these problems
in near future.

It appears more difficult (but not less important)
 to provide a supersymmetric generalization
of the formalism since, to our knowledge, a concept of a super-double
is not yet developed.
Because  the pair of a double and  of its maximally isotropic
subalgebra is nothing but a
classical limit
of the quasi-Hopf algebras \cite{D3} there appears a tantalizing possibility
of a relevance of the quasi-Hopf algebras  in a CFT description of
 the $D$-branes Poisson-Lie T-duality. It seems very probable
that  the duality   of the  quantum $D$-branes
is governed by the geometry and the representation theory of quantum groups.
We believe that many of the results obtained in past in the field
of quantum groups will find  a direct application in string theory in this
way.

\vskip1pc
\noindent {\bf Acknowledgement.} We thank A. Alekseev, L. \'Alvarez-Gaum\'e,
E. Kiritsis, C. Kounnas, M. Ro\v cek, R. Stora, A. Tseytlin, E. Verlinde
and E. Witten  for comments and discussions.

\end{document}